\documentclass{sig-alternate}

\usepackage{booktabs}
\usepackage{amsmath}
\usepackage{graphicx}
\usepackage{url}
\usepackage[normalem]{ulem}
\useunder{\uline}{\ul}{}
\usepackage{multirow}

\begin{document}
%

\title{Compiling Deep Learning Models for Custom Hardware Accelerators}


\numberofauthors{4}

\author{
\alignauthor
Andre Xian Ming Chang
       \affaddr{Purdue University Department of Electrical Engineering}\\
       \email{amingcha@purdue.edu}
\alignauthor 
Aliasger Zaidy
       \affaddr{Purdue University Department of Electrical Engineering}\\
       \email{azaidy@purdue.edu}
\alignauthor
Vinayak Gokhale
       \affaddr{Purdue University Department of Electrical Engineering}\\
       \email{vgokhale@purdue.edu}
\and  
\alignauthor 
Eugenio Culurciello
       \affaddr{Purdue University Department of Electrical Engineering}\\
       \email{euge@purdue.edu}
}

\maketitle
\begin{abstract}
Convolutional neural networks (CNNs) are the core of most state-of-the-art deep learning algorithms specialized for object detection and classification. CNNs are both computationally complex and embarrassingly parallel. Two properties that leave room for potential software and hardware optimizations for embedded systems. Given a programmable hardware accelerator with a CNN oriented custom instructions set, the compiler's task is to exploit the hardware's full potential, while abiding with the hardware constraints and maintaining generality to run different CNN models with varying workload properties. Snowflake is an efficient and scalable hardware accelerator implemented on programmable logic devices. It implements a control pipeline for a custom instruction set. The goal of this paper is to present Snowflake's compiler that generates machine level instructions from Torch7 model description files. The main software design points explored in this work are: model structure parsing, CNN workload breakdown, loop rearrangement for memory bandwidth optimizations and memory access balancing. The performance achieved by compiler generated instructions matches against hand optimized code for convolution layers. Generated instructions also efficiently execute AlexNet and ResNet18 inference on Snowflake. Snowflake with $256$ processing units was synthesized on Xilinx's Zynq XC7Z045 FPGA. At $250$\,MHz, AlexNet achieved in $93.6$\,fra\-mes/s and $1.2$\,GB/s of off-chip memory bandwidth, and $21.4$\,fra\-mes/s and $2.2$\,GB/s for ResNet18. Total on-chip power is $5$\,W.
\end{abstract}

\category{1.2}{Hardware for Embedded Systems}{HW/SW co-design for embedded systems}


\section{Introduction}
The deep learning field has grown in popularity in recent years with the success of back-propagation based models. For the past few years, CNNs have consecutively achieved the state-of-the-art accuracy for classification tasks \cite{alexnet_owt, googlenet, resnet} on large image datasets \cite{imagenet}. Those models are embarrassingly parallel, but exploiting parallelism for all existent models is a design problem with room for software and hardware exploration to achieve better performance per power. 

Snowflake \cite{snowflake} is a scalable and programmable, low-power accelerator for deep learning with a RISC based custom instruction set. Snowflake architecture was designed to provide high performance, given optimal sequence of instructions. But, manually crafting assembly like instructions can be cumbersome and error prone specially when a model is composed of several layers like in ResNet \cite{resnet}. Even if one was patient enough to manually write code for some of state-of-the-art deep learning models, further customization on both sides: on the hardware and software would require modifying thousands of lines of assembly code, preventing experimentation on custom system for deep learning. 

In this work, we present a compiler, which is responsible for generating instructions and managing data in the main memory. We designed a generic software structure to go from high level model representation from Torch7\cite{sw_torch7} down to an instruction stream that runs Snowflake. The main contributions of this work are:
\begin{itemize}  
\item A software framework to generate custom instructions for CNN targeted hardware accelerator.
\item Deciding whether to send maps data multiple times per set of kernels or send kernels multiple times per set of maps data for optimal bandwidth usage.
\item Communication load balancing to better utilize available memory bandwidth.
\end{itemize}

Snowflake was implemented on Xilinx's Zynq XC7Z045 FPGA \cite{xilinx:ZC706}. The system was benchmarked with AlexNet and ResNet18 \cite{alexnet_owt,resnet} pre-trained models. At $250$\,MHz, AlexNet achieved in $93.6$\,frames/s and $1.2$\,GB/s of off-chip memory bandwidth, and $21.4$\,frames/s and $2.2$\,GB/s for Res\-Net18. The following sections present background and related work, overview of Snowflake hardware architecture, overview of instruction set, details of compiler implementation and the obtained results.

\section{Background and related work}
Before designing a compiler for custom deep learning accelerators, lets briefly go through commonly used layers in CNN models that we will be targeting in this paper:

\textbf{Spatial Convolution (CONV)} is a $3$D convolution of an input volume with a group of $3$D kernels that result in extracted features from the input. Each $3$D kernel is associated with a bias value. Number channels is the z-axis of input volume. Input volume is called maps and a slice of it is a map. A window is the size of one kernel convolved with part of input volume. Operations in a compute window are independent, hence it is well suited to multi-core processors: GPUs \cite{gpucnn} and other designs using ASIC \cite{chen_eyeriss_isca,cavigelli_origami} and FPGAs \cite{nnx,zhang_cnnfpga}.

\textbf{Activation unit} is a non-linear function that some layer's outputs go through. Some examples are: rectified linear unit (ReLU), tanh and sigmoid. In this work, we only use ReLU. 

\textbf{Max pooling (Maxpool)} is a down-sampling technique to achieve data invariance and to compress feature representation. Max pooling is element-wise comparison and its result is the highest value in a processing window. 

\textbf{Average pooling (Avgpool)} is similar to Max pooling, but instead of getting the highest value in a window, it averages out its values. Average pooling can be implemented as a CONV with a single weight value of inverse of window size. Multiplying and accumulating all values in a window with this weight gives the average value of a window.

\textbf{Residual addition} or bypass is used in ResNet models \cite{resnet}. The output values of a CONV are element-wise added with a previous layer's input. In hardware, we want to add those bypass values as output results are being produced by a CONV layer to save communication cost. Thus we need to keep track of previous input layers and to conditionally issue an extra instruction.

\textbf{Fully connected (FC) layers} are used to map the features into a classification vector that has the "final answer" of the network and it is usually the last layer of a CNN model. FC layer is a data movement intensive operation because it provides limited data reuse. Thus, memory bandwidth is a bottleneck for running FC layers. Weight compression and weight pruning based on network sparsity are techniques that lower memory bandwidth requirement for this type of workload \cite{han_eie, iccad_compression}.

\cite{farabet_neuflow} presents a compiler for a custom CNN accelerator using Torch5 models. Their approach is to map Torch5 models into a set of pre-defined sequence of control signals for DMA transfers and processing units. Custom hardware and instruction generation software was developed for Caffe \cite{sw_caffe} in \cite{zhang_caffeine,hw_neurostream}. This compiler is the first to generate to custom instructions for hardware accelerator from Torch7 \cite{sw_torch7} or Pytorch models. 

The system in \cite{zhang_caffeine} maps FC layers and CONV layers into a uniformed control representation: input or weight major processing. This work also uses uniform representation, but with a finer granularity defined as trace, which is any contiguous sequence of multiply and accumulate. This allows finer hardware and algorithmic optimizations. 

Memory transfer friendly computation tiling for CNN accelerators was explored in \cite{memory_cnn, striptilecnn}. In \cite{hw_neurostream}, block tiling with x-y axis ordering was used. They store tiles with extra overlap regions, called augmented-tiles, in DRAM to avoid multiple DMA transactions. This work also stores overlapped regions but it tiles at the granularity of row strips with channel major ordering to lower overlapped data replication. This lowers the required memory bandwidth.

Other domain-specific instructions set for CNN were presented in \cite{isa_cambricon}, can be added into Snowflake's compiler, because they also use vector compute instructions and scratchpad on-chip memory loads instructions. The intermediate representation and the techniques presented can be also be integrated into conventional frameworks \cite{sw_llvm}, which is left for future work. 


\section{Snowflake hardware overview}
\begin{figure*}
\centering
  \includegraphics[width=6.4in,height=2.2in]{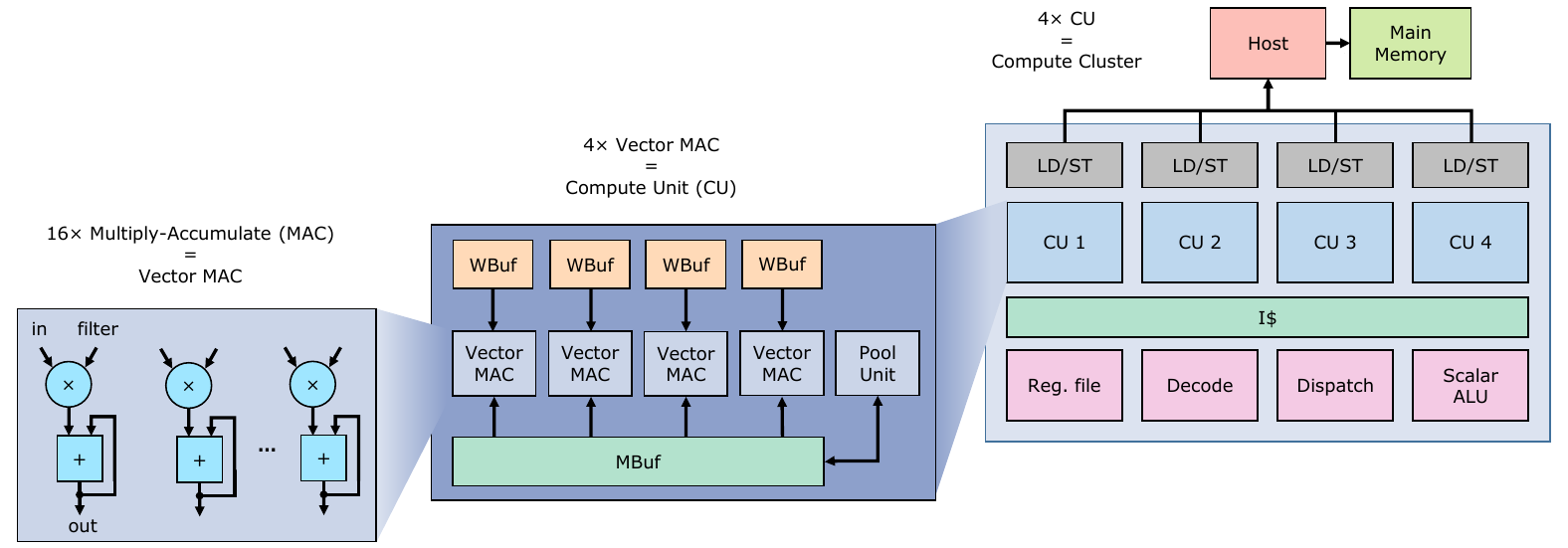}
  \caption{Snowflake architecture block diagram. On the left, a vMAC is a group of MAC units. In the center, a CU is composed of a group of vMACs with data buffers. In the right, a group of CUs forms a compute cluster that shares the control unit and load/store units.}
  \label{fig:snowflake}
\end{figure*}

Snowflake was presented in \cite{snowflake}. This section summarizes the main hardware concepts that will be needed to develop a compiler. For readers' convenience Snowflake's diagram is shown in figure \ref{fig:snowflake}. 

The main building block of Snowflake's convolution engine are $16$\,bit multiply and accumulate units (MACs). A vector MAC (vMAC) is comprised of $16$ MACs, that process $256$\,bits in one cycle. A compute unit (CU) is composed of $4$ vMACs. Each vMAC has a private kernel scratchpad buffer (WBuf) and every vMAC in a CU shares the input data through the maps scratchpad buffer (MBuf). Data transfer time is overlapped by MAC compute time by using double buffer strategy.

Numerous CUs can be grouped into compute clusters. A compute cluster has a control unit, which is a RISC based pipeline. There are $4$ load/store units that access the host main memory through DMA using AXI protocol. Vector comparator units (Pool Unit) are used for max-pool operations. RISC based instructions are loaded into the instruction cache (I\$). The synthesized Snowflake was aimed for embedded system workloads, thus only one cluster with $4$ CUs was instantiated. Each maps bank has $64$\,KB and each vMACs weight buffer is $8$\,KB. Instruction cache is $4$\,KB. Two ARM Cortex-A9 CPUs function as the host processor for the Snowflake implementation. Snowflake is clocked at $250$\,MHz and the ARM CPUs are at $666$\,MHz.

\subsection{Control pipeline}
The control pipeline provides the CUs control signals given a stream of instructions from the instruction cache. It is a pipeline with $5$ stages: fetch, decode, dispatch, execute and register writeback.

In the \textbf{fetch stage}, instructions are read from the instruction cache. The instructions are $32$\,bit. The \textbf{decode stage} turns fetched instructions into signals for operation modes, sign extends immediate values and gives addresses to access the register file. Decode stage also performs true dependency or read-after-write (RAW) hazards detection. RAW causes decode stalls, which is a pipeline stall but not necessarily a CU stall. The \textbf{dispatch stage} is responsible for identifying the resources needed for the current instruction and issuing register file read. Snowflake has $32$ $32$\,bit registers. If the decoded instruction is a vector operation, then CUs receive signals to fetch data and start processing a vector. Scalar computations are implemented in execute stage. The \textbf{execute stage} gets the dispatch signals and starts the compute resources for a scalar operation. Scalar unit is composed of multiplier, adder and comparator. Previous stages are single cycle latency, whereas the execute stage is $2$ cycle. Finally, the \textbf{register write-back stage} writes scalar results to the register file.




\section{Custom instruction set}\label{sec:isa}
Snowflake's instruction set contains $13$ instructions: MOV, MOVI, ADD, ADDI, MUL, MULI, MAC, MAX, VMOV, BLE, BGT, BEQ and LD. The $32$\,bit instructions are grouped into four categories: data movement, compute, flow control and memory access. Most instructions have in common four properties: $4$\,bit operand code, $1$\,bit mode select, $5$\,bit register selects (destination and two source registers) and a immediate field.

\textbf{Data movement}: Instruction MOVI moves a $23$\,bit immediate value into a register. MOV moves data between registers with a optional $5$\,bit left shift. VMOV is a vector move from buffer into compute units. It fetches a buffer block and load it into an operand register of a compute unit defined by select. VMOV is used to load the CONV/FC layer's bias into the MACs or to load the bypass values in residual add for ResNet.


\textbf{Compute}: ADD is a simple register to register add and ADDI is register with immediate add. MUL and MULI are register to register and register to immediate multiply, respectively. The vector compute units are controlled by MAC/MAX instructions. MAC multiplies and accumulates from contiguous sequence of data (trace) in maps and weights buffers. MAX is another vector instruction that has similar behavior to MAC. It performs comparisons with a retained previous vector. After a vector instruction finishes, a store to buffer or store to main memory is issued, thus there is not an explicit store instruction. 

Snowflake's MAC instruction has two modes of operations: cooperative (COOP) and independent (INDP). In COOP mode, all MACs in one vMAC work together to produce one value of the output map. Each MAC processes a different channel of one kernel, and the results of all $16$ MACs are added together by an extra adder called gather adder to produce one value. In independent mode, all MACs in one vMAC work independently on different kernels and map values are broadcast to produce $16$ different output map values. More details on custom instructions are presented in \cite{snowflake}.

\textbf{Flow control}: BLE, BEQ, BGT are branch instructions that compare the value at Rs1 to the value at Rs2. The immediate is the PC offset when the condition is true. Branches take $4$ cycles to go through the pipeline, thus it leaves $4$ branch delay slots to be filled with instructions. Only one pair of true RAW dependent instructions is allowed in the branch delay slots.

\textbf{Memory access}: LD instruction loads data from main memory to one of Snowflake's buffers. Snowflake can have multiple load units that can start independent load streams, within the off-chip memory bandwidth constraints. LD have select modes that allow a processing choice of weights broadcast and different maps, or maps broadcast with different weights.

\section{Methodology}\label{sec:snowb}
Given the custom hardware constraints, the compiler is responsible to orchestrate the hardware for all sorts of layers. How to load balance the communication ports for better bandwidth utilization, how to partition maps and kernels to fit in on-chip buffers, how to issue compute and load instructions without stalling and how to balance between loops and unrolled instructions are some software design decisions.

The compiler performs three major tasks: parse high level representation of a model, instruction generation and instruction deployment into hardware. Each task has steps that are described in following subsections. 


\subsection{Model parsing}\label{sec:parse}
In this task, the start point is a model representation from Torch7, and the end point is a data structure that contains all the information to easily generate Snowflake instructions. There are $5$ steps to reach our goal.





\emph{Thnets}\footnote{\url{https://github.com/mvitez/thnets}} is an open-source library that provides means to read a Torch7 model representation file and to convert it into a C data structure. Using Thnets' functions, \textbf{step\,1} loads the parameters of each layer in the model into a layer object. This step scans through all layers and ignores non-sequential inter-layer relations that happens in parallel layer paths. The layer objects are serialized into a doubly linked list. Snowflake will process each element in the list in sequence.

In some models, such as GoogLe\-Net \cite{googlenet} and ResNet \cite{resnet}, not all the layers are sequential. Some layers share their input and output, thus some layers in the list are labeled according to their parallel path. \textbf{Step\,2} scans the model to get the inter-layer relations, and creates a dependency label for each layer object. This label indicates whether the layer is only connected to its previous and next layers or not. This label translates into how each layer share their maps data in pre-allocated main memory regions.


\textbf{Step\,3} processes each layer's information and its neighboring layers to decide how to decompose and generate instructions for different layers. Snowflake hardware parameter object is globally shared among functions to create hardware dependent structures: maps tile, kernel tile and load objects. The main hardware constraints are: 
\begin{itemize}
\item Instruction cache size: Snowflake instruction cache is double banked with $512$ instructions per bank. But branching across instruction banks is not permitted. This affects how loops are broken down.

\item Data buffer size: this defines how to decompose the maps and weights into tiles. It also affects the required memory bandwidth, since smaller tiles means more overlapped data is loaded more frequently.

\item Memory bandwidth: computation stalls happen when required data has not arrived into the buffer, which is caused by the memory bandwidth constraint. It affects whether to loop kernels or maps. This will be explained in section \ref{ssec:kmloop}.

\item Instruction latency: vector instructions require variable latency to produce a result. Within these cycles we want to hide all other necessary operations: loop control, conditional branches, buffer address increment and load instructions. Another reason why a CU can stall is because there was not enough MAC/MAX instructions issued in sequence, not leaving enough cycles to overlap with bookkeeping instructions. 
\end{itemize}

Based on these constraints and the layer parameters, the compiler sets decision variables that chooses mode (COOP or INDP) to use, chooses loop breakdown based on instruction cache size, sets tile size limit based on data buffer size, chooses whether is better to fix map data and loop through kernels or vice-versa based on memory bandwidth constraint. 

\textbf{Step\,4} breaks down the layers' maps and weights data into tiles that fits into the data buffer. The maps are decomposed in tiles with output row granularity, meaning that each tile produces output row(s). Weights are decomposed in tiles with single kernel granularity. Single kernel size is input channels times window size. Based on the decision made in step $2$ and the neighboring layers, the tiles can have different data sizes. For instance, if we broadcast weights, then each CU works on different map tiles and the maps need to be decomposed such that all CUs will have the same amount of work. Inevitably, some remaining tiles won't be big enough to share among all CUs. Then some CUs must be disabled. Another example is in the context of ResNet where a CONV followed by a bypass needs two input maps: one for the CONV and one for the bypass. This special CONV needs to use both maps buffer banks simultaneously. Double buffering is done by using different buffer regions.


After creating a list of tiles to process, each element of the list will create operation lists in \textbf{step\,5}. For each tile, there are two major operations: load and compute. A load list carries information necessary for a load instruction: stream length, memory address and buffer address. Each load object is associated with a different tile. The load list is created taking into account subsequent tiles loads, so that Snowflake can process a tile while loading data for the next one. 

Compute objects contain information about a set of vector compute operations in a window, the number repetition in x and y directions and x and y offsets. This allows grouping of striding windows with same window size into one object. For example, multiple compute objects are needed for CONVs with padding, whereas a single object suffices in a CONV without padding. Compute objects will be translated into nested loops of MAC/MAXP instructions. The loop boundaries are the repeat variable and the vector instruction read buffer address is incremented by the offset variable. The compute object also has an extension with variables for VMOV if the layer is a CONV with bypass.

\begin{figure}[ht]
  \centering
  \includegraphics[width=3.1in,height=3.0in]{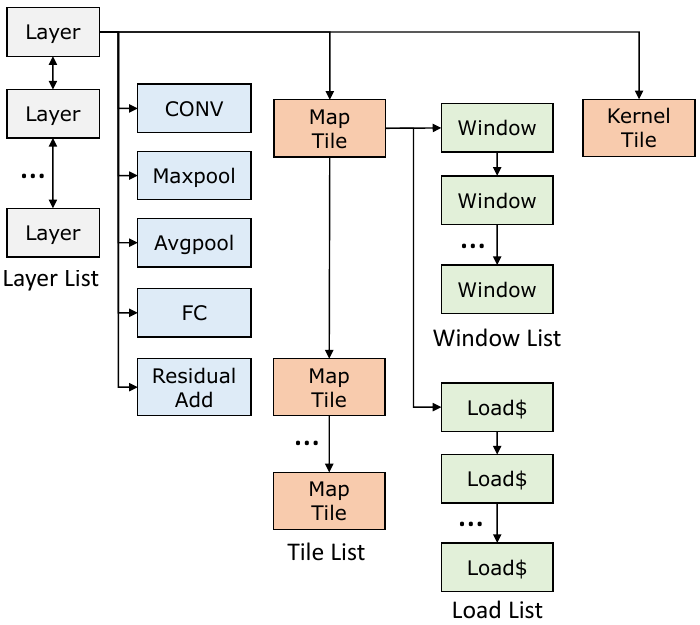}
  \caption{Example of the compiler's data structure. The arrows represent pointers. Objects are grouped into lists. Head and tail pointers of the lists are omitted.}
  \label{fig:datastruct}
\end{figure}

An example of Snowflake's data structure is shown in figure\,\ref{fig:datastruct}. A layer object can be one of the layer types: CONV, Maxpool, Avgpool, FC and Residual add. A layer has two lists of tiles: one for kernel and one for maps. Each tile object can instantiate a list of windows and a list of loads. A kernel tile does not have window list because compute operations are defined by maps tiles. Load list is in the tile list that is not being repeated in a loop. In the example shown in figure\,\ref{fig:datastruct}, all kernels are looped through each maps tile. Kernel tile object contains repeat variable that defines the kernel loop boundaries. Hence, there is one kernel tile object. Kernel loads are implicit in a repeating tile object parameters, thus there is not a load list in the kernel tile. If maps are being repeated in loop for each different kernel tile, then kernel tile objects have load list and maps wont.


\subsection{Instruction generation}
After the model parsing task completes, the compiler goes through the lists and create instructions accordingly. Each tile object creates a block of instructions that will be concatenated with other tile objects instructions. The compiler inserts load for the following instruction cache bank at the beginning of each instruction block and inserts a jump to next instruction bank at the end to allow instruction double buffering. Hence, before generating instructions, we need to predict how many instructions are needed for a tile and check if it fits into the instruction cache constraint, so that we can correctly insert instruction loads and jumps. The prediction step passes through all objects and creates a temporary instruction block for each tile. If the predicted number of instruction in a block is larger than instruction cache bank size, then different instruction generation strategy must be used for that tile. After an instruction count profile for each tile is generated, the program knows where it can safely insert an instruction load for next bank. Then the permanent instruction stream is created.

Most of the instruction stream structure was sketched in the data structure as a result from model parsing in section \ref{sec:parse}. There are three main goals that instructions blocks need to accomplish:

\begin{itemize}
\item Initialization or register reset: the first tile of the model needs to populate all buffers in Snowflake creating a initialization latency. The following tiles needs to reset some loop control registers and set the reserved registers to the correct output locations.
\item Compute data: each tile has their compute list, which defines a maximum of $3$ nested loops. The inner loop is to accumulate traces, the second loop is stride along x-axis and the outer loop is stride along y-axis. Respectively, the loop limits are defined by kernel height, repeat x variable and repeat y variable. 
\item Load data: buffers needs to be updated for each compute tile, so that the following compute section can proceed without stalling for data. Both weights buffer and maps buffer loads need to be inserted in between compute operations, so that data coherence is maintained. 
\end{itemize}

\begin{figure*}[ht]
\centering
  \includegraphics[width=6.4in,height=2.2in]{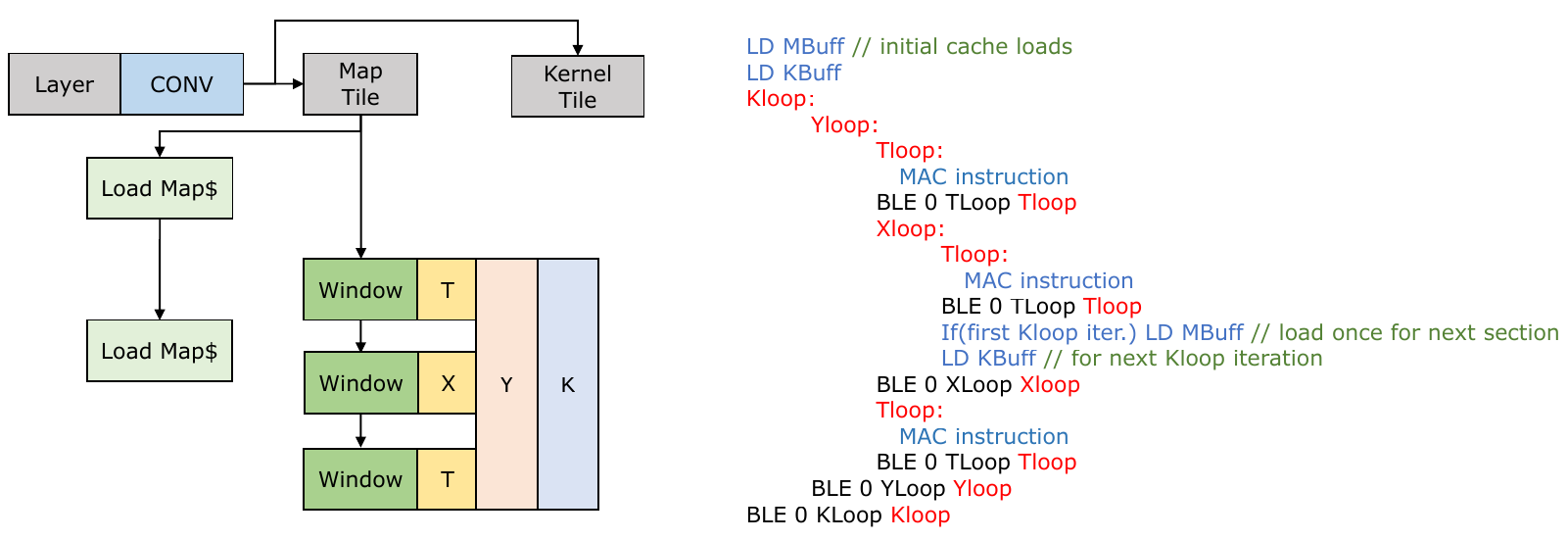}
  \caption{Example of instruction generation for a CONV layer. Different layers create different number of objects, but they all become MAC/MAX for window objects and LD for load objects.}
  \label{fig:gen_instr}
\end{figure*}

Figure\,\ref{fig:gen_instr} shows an example of instructions for a CONV tile, where T denotes trace loop, X x-axis loop, Y y-axis loop and K kernel loop (Kloop). The compiler first generates instructions for the initial map and weight buffer loads with the parameters in the first load object. Then it creates a kernel loop based on the kernel tile object. If the map tile window objects have non-unity repeat Y then a Y loop is created. Inside Y loop, T and X loops are created depending on the window objects variables. The load for the next tile, which is the second load MBuf object, is inserted in the X loop. The load insertion is needed because the compiler needs to guarantee that previously issued vector instructions have finished using a buffer bank before issuing new data load to that buffer bank. One way is to issue $16$ vector instructions that will fill the trace buffer with new vector instructions, which guarantees that there no old vector instructions pending. 

Depending on the CONV case, some of the loops are removed. For example, $1\times1$ CONVs don't have trace loop. Depending on bandwidth constraint, map and kernel tile objects can be generated, such that the kernel loop can become map loop, which means maps are sent multiple times for one tile of kernel. 

Window objects can be broken down into multiple loops, which is based on the instruction latency versus instruction size trade-off. We can only put limited amount of bookkeeping work in between two consecutive vector instructions. If those operations takes on average more cycles than MAC/MAX latency then CUs would stall because there is not enough MAC/MAX instruction being issued in sequence. A CONV with higher MAC latency allows more freedom to add flow control instructions, making the instruction stream more generic and with fewer instructions. On the other hand, $1\times1$ CONVs have lower MAC latency, which restricts the number of bookkeeping instructions that can be overlapped with the MAC instruction's latency. For example, if a MAC takes $16$ cycles to finish, then having $20$ instructions (loads, branches and other bookkeeping) in between MAC instructions will stall the CUs. In this case, breaking the loops or full loop unrolling will reduce the amount of bookkeeping instructions between consecutive instructions. But this increases the instruction count.  One extreme is that all loops are completely unrolled. 




Window objects also have the buffer address of bias value associated with each output map produced by a kernel, so a VMOV is needed before each Kloop iteration. For a residual add, a VMOV for each write-back MAC instruction is necessary because output value is added with a bypass value. Note that only the last MAC instruction of a trace loop is a write-back MAC instruction. VMOV also adds additional bookkeeping instructions for incrementing and resetting addresses and word select. This causes issue in extreme cases when there are not enough MAC instructions to hide the bookkeeping operations, like in the last $1\times1$ CONVs of ResNet18 and ResNet50.

Another part is that load instruction can become comparably large. For example, if weights are not broadcast, then there will be a load for each weights buffer. In a $4$ CU system, there will be $16$ weight LDs plus load ID bookkeeping operations. For a low MAC/MAX latency CONV case, the $16$ loads should be spread out and interleaved with the MAC/MAX instructions. Load unit balance is also necessary to better utilize the available memory bandwidth. It is better to break a single large load transaction into multiple smaller loads to prevent the CUs from stalling for incoming data.


Instruction granularity optimization: register assignment, branch delay slot filling and instruction reordering are topics for future 
work. For this paper, register assignment is statically defined to avoid unnecessary register saving instructions. Branch delay slots are filled and instruction order is manually optimized for a small subset of the main tasks: compute and load. Finding the sweet spot between fully handwritten code and generic pieces of optimized instructions that achieves high-performance for most use cases is up to further study.
\vfill\null

\subsection{Instruction deployment}
Last task is to run Snowflake. This task loads and arranges weights and biases data from a trained model. The weights and bias need to be arranged differently based on the workload break down and the compute decision made earlier. For instance, each vMAC works on a different kernel in COOP mode, whereas in INDP mode each vMAC works on $16$ different kernels, hence in INDP we need to group $16$ kernels in one vMAC weight buffer. Instruction deployment task also need to load the input image for the first layer of the model.

Snowflake uses CMA (Contiguous Memory Allocator) for memory to FPGA communication. All data need to be placed into CMA allocated region of memory. Different regions in CMA are allocated according to layer dependencies (step $2$ in section \ref{sec:parse}). Different regions are allocated for each layer's weights. After that, some configuration registers enables an initial load instruction to populate Snowflake's instruction cache with the first set of instructions. The software polls an output counter register to check whether processing has finished or not. Finally, for validation purposes, we wrote a software implementation of the model's layers using Q8.8 to simulate Snowflake's compute operations. Result checking allows layer by layer validation.

The data representation of choice for hardware and software was Q8.8, which has been shown to have insignificant accuracy degradation compared to neural networks implemented in $32$\,bit floating point \cite{fixpoint}. Nevertheless, other number representations can be used in the system. Pre-trained ResNet18 \footnote{\url{https://github.com/facebook/fb.resnet.torch}} was profiled on ImageNet dataset \cite{imagenet} using Q8.8 and Q5.11 fixed point precisions. Top-5 accuracy using $32$\,bit float was $89$\% , Q8.8 was $84$\% and Q5.11 was $88\%$.

\section{Results}
We compared hand optimized instruction stream versus code generated from Snowflake's compiler. Performance results for AlexNet, ResNet18 and ResNet50 model were measured. Finally, this section presents a discussion of techniques that were used: kernel or maps data loop and communication load balance.

\subsection{Generated instructions}
Using all the techniques described in previous section, the compiler generates an instruction stream that achieves performance comparable to handcrafted instructions as shown in Table \ref{tab:hand}. In the table, CONVs parameters are, respectively, input size, kernel size, input plane, output plane, stride and padding. \emph{Auto} stands for compiler generated code and \emph{hand} is handwritten code. Auto-generated code has higher instruction count ($437$ more), but it achieves similar execution time compared to hand optimized code, which exploits manual optimizations such as filling branch delay slots and instruction reordering. We have only compared some AlexNet layers because models in handwritten instruction are human error prone and tedious. The results for auto-generated instruction for models are shown in table\,\ref{tab:models}. $224\times224$ images was used as model's input.

\begin{table}[h]
\centering
\caption{Hand optimized code (hand) versus auto-generated instructions (auto).}
\label{tab:hand}
\begin{tabular}{|l|l|r|}
\hline
Layer & Code & Time {[}ms{]} \\ \hline
\multirow{2}{*}{27x27,5x5,64,192,1,2} & Hand & 3.256 \\ \cline{2-3} 
 & \textbf{Auto} & 3.261  \\ \hline
\multirow{2}{*}{13x13,3x3,192,384,1,1} & Hand & 1.627  \\ \cline{2-3} 
 & \textbf{Auto} & 1.624  \\ \hline
\multirow{2}{*}{13x13,3x3,384,256,1,1} & Hand & 2.188\\ \cline{2-3} 
 & \textbf{Auto} & 2.187 \\ \hline
\multirow{2}{*}{13x13,3x3,256,256,1,1} & Hand & 1.462  \\ \cline{2-3} 
 & \textbf{Auto} & 1.458 \\ \hline
\end{tabular}
\end{table}


The compiler results show its main contributions: to provide means to test models, ensure output correctness and to allow further exploration. Some inefficiency is caused by cold buffer misses, memory bandwidth limitation and non-overlapped Maxpool layers. Execution time for all models does not account for FC layer times, since FC layers are inherently bandwidth limited operations. Those issues will be addressed in future hardware and software developments.

\begin{table}[h]
\centering
\caption{Results for models using Snowflake's compiler}
\label{tab:models}
\begin{tabular}{|l|r|r|}
\hline
 Model & Exec. Time {[}ms{]} & BW {[}GB/s{]} \\\hline
AlexNetOWT & 10.68 & 1.22 \\\hline
ResNet18 & 46.77 & 2.25 \\\hline
ResNet50 & 218.61 & 1.87 \\\hline
\end{tabular}
\end{table}

\subsection{Loop rearrangement for bandwidth constraints}\label{ssec:kmloop}
Unlike GPUs and ASIC designs, FPGA accelerators are limited mostly by their off-chip memory bandwidth. While a GPU's optimized memory interconnects can achieve a high bandwidth of $112$\,GB/s \cite{gpunvidia}, Xilinx ZC706 board \cite{xilinx:ZC706} can achieve $4.2$\,GB/s bi-directional bandwidth with the AXI ports \cite{xilinx:axis4lite}.

\begin{figure}[!t]
\centering
\includegraphics[width=3.3in,height=3.3in]{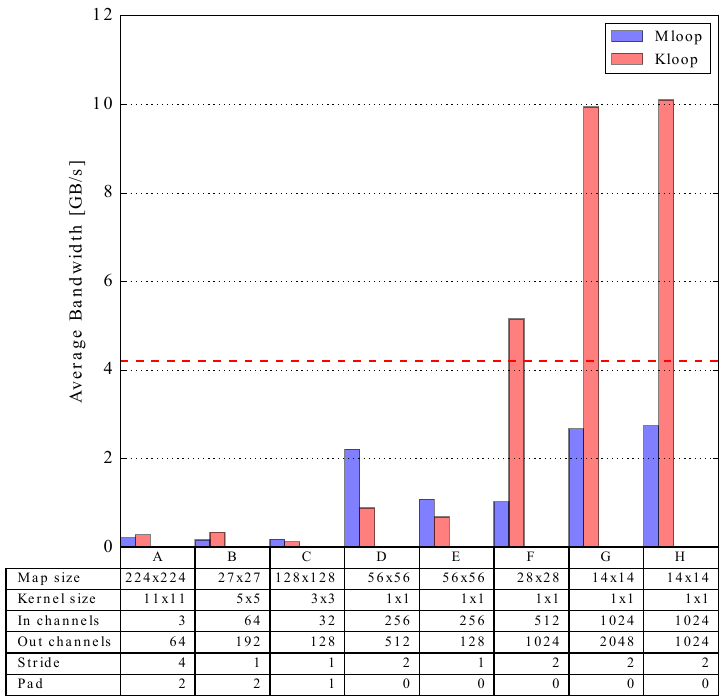}
\caption{Required memory bandwidth in Mloop or Kloop mode for various CONV examples.}
\label{fig:mkloop}
\end{figure}

Loop rearrangement is a method that reduces total amount of data movement between main memory and hardware accelerator leading to memory bandwidth savings. Some CONV layers have large kernels, whereas others have large maps, but usually neither completely fits into the buffer. Maps and kernels need to be partitioned and processed in buffer sized tiles. A map tile needs to go through each kernel tile, leading to repeated kernel loads when the next map tile is loaded. Alternatively, a kernel tile needs to be processed with every map tile, resulting in repeated map loads for the following kernel tile. The total amount of data moved is different depending on kernel/map load repetition for a particular CONV layer. Figure\,\ref{fig:mkloop} shows some examples of CONVs that have lower bandwidth requirement with maps load repetition and vice-versa. Mloop is abbreviation for repeated maps data and Kloop is abbreviation for repeated kernel data. A red dashed line indicates the memory bandwidth limit of the development board. CONVs A and B are from AlexNet model. Their required memory bandwidth is below the limit, so choosing between Mloop or Kloop wont significantly affect performance. CONVs G and H are examples from Resnet50 model and their required memory bandwidth is above the limit for the Mloop mode. Kloop mode is necessary for those layers.

\subsection{Communication load balance}
Snowflake has $4$ load/store units, and properly distributing LD instructions to all units prevents CU stalls due to data transfer. Issuing a single map load to a unit while distributing kernels for all units will lead to unbalanced load units workload. A better approach is to break the maps data into multiple load instructions and distribute evenly with the kernel loads. Communication load balancing optimizes the load unit usage, and thus results in better usage of available memory bandwidth. Load imbalance is a measure of how evenly data is distributed. The percent imbalance metric equation \ref{eq:ldbal} is commonly used \cite{loadimbalance}. Where $L_{max}$ is maximum load for any load unit and $\mu_{L}$ is the mean load over all load units. 

\begin{equation} \label{eq:ldbal}
C_{L} = (\frac{L_{max}}{\mu_{L}}-1)\times100\%
\end{equation}

\begin{table}[h]
\centering
\caption{Speed up versus load imbalance.}
\label{tab:ldbal}
\begin{tabular}{|l|l|}
\hline
Load Balance {[}\%{]} & Speed up \\ \hline
5 & 1.658 \\ \hline
17 & 1.656 \\ \hline
42 & 1.652 \\ \hline
102 & 1.644 \\ \hline
114 & 1.297 \\ \hline
132 & 1.000 \\ \hline
\end{tabular}
\end{table}

Table\,\ref{tab:ldbal} shows the speedup achieved by reducing the load imbalance on a CONV $1\times1$ with $1024$ input channels, $2048$ output channels and stride $2$. The load imbalance percent is measured and averaged out for all tiles. The worst imbalance in the table\,\ref{tab:ldbal} is the case when kernel and maps uses two load units. The measured speedup in the execution time is compared with the worst load imbalance. This shows that finer load balancing has gains in performance up to a certain limit when data loads are mostly overlapped with vector computation. From this point, reducing memory latency by better load distribution results in diminishing improvements.

\section{Conclusions}
This work presented a complete software design flow from high level model definitions created with popular deep learning tools (Torch7) down to custom architecture for accelerating deep learning. This compiler was implemented to provide hardware usability, while efficiently utilizing all hardware resources for various CNN workloads. This work addresses software design points, such as model structure parsing, workload breakdown, loop rearrangement and memory access balancing. Those techniques were tested on the Snowflake custom accelerator, but they can be applied to other custom accelerators.

\bibliographystyle{abbrv}
\bibliography{all}  

\end{document}